\documentclass[aps,prl,showpacs,twocolumn,amsmath,amssymb]{revtex4}
\usepackage[]{graphicx}
\usepackage{amsmath}

\begin{document}

\title{Direct observation of mode-coupling instability in two-dimensional plasma crystals}

\author{L. Cou\"edel}
\email{lcouedel@mpe.mpg.de}
\author{V. Nosenko}
\author{A. V. Ivlev}
\author{S. K. Zhdanov}
\author{H. M. Thomas}
\author{G. E. Morfill}

\affiliation{Max Planck Institute for Extraterrestrial Physics, 85741 Garching, Germany}

\begin{abstract}
Dedicated experiments on melting of 2D plasma crystals were carried out. The melting was always accompanied by spontaneous
growth of the particle kinetic energy, suggesting a universal plasma-driven mechanism underlying the process. By measuring
three principal dust-lattice (DL) wave modes simultaneously, it is unambiguously demonstrated that the melting occurs due to
the resonance coupling between two of the DL modes. The variation of the wave modes with the experimental conditions,
including the emergence of the resonant (hybrid) branch, reveals exceptionally good agreement with the theory of
mode-coupling instability.
\end{abstract}


\maketitle

Strongly coupled complex (dusty) plasmas play an important role in the existing hierarchy of soft matter states
\cite{Morfill09}. Along with complex fluids and granular media as prominent examples of ``regular'' soft matter, complex
plasmas represent natural model systems which enable remarkably broad interdisciplinary research. The characteristic length-
and timescales in such systems are dramatically stretched (e.g., in complex plasmas -- to hundreds of microns and tens of
milliseconds, respectively). Therefore, numerous generic processes occurring in classical fluids or solids can be studied in
greatest detail, at the most fundamental ``atomistic'' level \cite{Morfill09,Fortov05,VladimirovBook}. Especially important
here are 2D systems, where the complete information about all particles can be obtained at each moment of time. The current
experimental capabilities make 2D complex plasmas ideal for comprehensive experimental studies of the atomistic dynamics
\cite{Morfill09,Fortov05,Thomas96,Chiang96}.

The investigation of atomistic processes that trigger the melting and crystallization is of particular interest. It is well
known that the mechanisms resulting in the destruction of the long-range crystalline order in 2D systems can be very
different from those operating in 3D crystals: The classical Kosterlitz-Thouless-Halperin-Nelson-Young (KTHNY) theory of 2D
melting \cite{KTHNY} predicts two consecutive phase transitions (with an intermediate, so called ``hexatic phase'' in
between). The alternative theory \cite{Chui} relates 2D melting with the formation of dislocation chains (``grain
boundaries'') percolating the system. These are generic melting mechanisms which can operate in very different strongly
coupled systems \cite{Murray87,Zahn00,Quinn96,Sheridan08,Nosenko09}.

When studying generic classical phenomena occurring in regular liquids and solids, the {\it relevance} of a model system (be
it colloidal suspensions, granular media, or complex plasmas) becomes crucial. Of course, careful analysis is required in
the context of a given phenomenon (or, class of phenomena), but one can certainly identify essential common principles. In
particular, the applicability of the Hamiltonial approach for the analysis of atomistic dynamics (for instance, to
investigate kinetics of phase transitions) is one of these basic principles. This implies that non-Hamiltonian
(non-conservative) behavior peculiar to model systems should play minor role at the relevant timescales \cite{footnote1}.

Non-Hamiltonial behavior of particle ensembles is often a manifestation of asymmetry, or {\it nonreciprocity}, in the
interparticle interactions. Such asymmetry can be considered as a result of the system openness: In the presence of strong
flow of the ambient plasma the screening cloud around each charged particle becomes highly asymmetric. These clouds, which
are usually referred to as ``plasma wakes'' \cite{Vladimirov95,Vladimirov96}, play the role of an (external) ``third body''
in the interparticle interaction and hence make it nonreciprocal \cite{Melzer99,footnote2}. One should point out that such
nonreciprocity is quite universal and can be very important for the hydrodynamic interactions of colloids \cite{Araki08} or
droplets in conventional fluid flows \cite{Beatus06}.


The nonreciprocity of pair interaction in complex plasmas provides a very efficient mechanism of converting the energy of
the flowing ions into the kinetic energy of microparticles \cite{Morfill09,Fortov05,Melzer96}. In particular, this can be
associated with dust-lattice (DL) wave modes \cite{Fortov05,Vladimirov97,Couedel09}: The kinetic energy can grow due to
resonant particle-wake interactions when different oscillation modes of interacting particles are properly synchronized.
Ivlev and Morfill \cite{Ivlev01} suggested that the resonance between DL wave modes in a 2D plasma crystal can trigger the
{\it mode-coupling instability} which, in turn, can cause the crystal melting. By comparing the melting experiments with the
molecular dynamic simulations, this hypothesis has received strong confirmation \cite{Ivlev03}. However, the direct proof
that the crystalline monolayers can indeed melt due to the mode-coupling instability (and even that the instability itself
can be observed in experiments) was still outstanding. One should point out that the other known plasma-related scenarios of
2D crystal melting either require the presence of localized defects -- single particles (or even sublayers) above or below
the monolayer \cite{Schweigert00}, or they are associated with the charge fluctuations and can only operate at very low
pressures (well below 1 Pa) \cite{Vaulina99,Nunomura99}.

In this Letter we report on dedicated experimental investigation of 2D plasma crystal melting. We explored a fairly broad
range of experimental parameters chosen well inside the crystalline regime predicted by the generic (KTHNY or
grain-boundary) melting scenarios \cite{Nosenko09}. The melting was always accompanied by a spontaneous growth of the
microparticle kinetic energy, which clearly suggests a universal plasma-driven mechanism underlying the process. By
implementing simultaneous measurements of the dispersion relations for three principal DL wave modes, we unambiguously
demonstrated that the melting occurred according to the mechanism predicted by Ivlev and Morfill. The variation of the DL
modes with the experimental conditions, including the emergence of the resonant (hybrid) mode, revealed exceptionally good
agreement with the theoretical predictions of the mode-coupling instability.

\begin{figure}
\includegraphics[width=6.0cm,clip=]{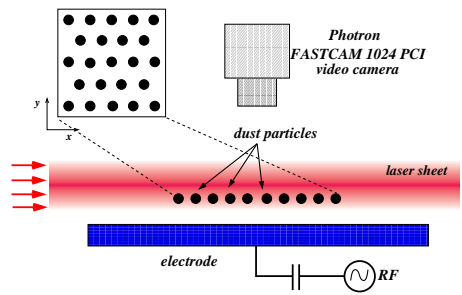}
\caption {\label{fig1} Sketch of the experimental setup. Microparticles are confined above the RF electrode and are illuminated
with a horizontal laser sheet having a Gaussian profile in the vertical direction (with a standard deviation $\sigma=75~\mu$m).
The monolayer is levitated well below the peak of the laser intensity, which results in strong intensity variations of the
scattered light upon the vertical displacement of individual particles. The upward/downward displacement corresponds to
positive/negative intensity variation. Microparticles are recorded from top at a speed of 250 frames per second.}
\end{figure}

The melting was systematically investigated in experiments performed with a (modified) GEC chamber, in a capacitively
coupled rf glow discharge at 13.56 MHz (see Fig.~\ref{fig1}). The argon pressure was between 0.4 Pa and 1 Pa and the forward
rf power was between 5 W and 20 W (which corresponded to rf peak-to-peak voltage between 175 V and 310 V).
A horizontal monolayer was formed by levitating melamine-formaldehyde particles in the plasma sheath above the lower rf
electrode. We used particles of two different sizes, $8.77\pm0.14~\mu$m and $9.15\pm0.14~\mu$m diameter. A high quality of
the monolayer (with no detected particles above or below) was ensured by using
a side-view camera. The diameter of the obtained crystalline structure was up to $\simeq60$~mm, depending on the number of
injected particles (we explored a range from 1000 to 15,000).



\begin{figure}
\includegraphics[width=6.3cm,clip=]{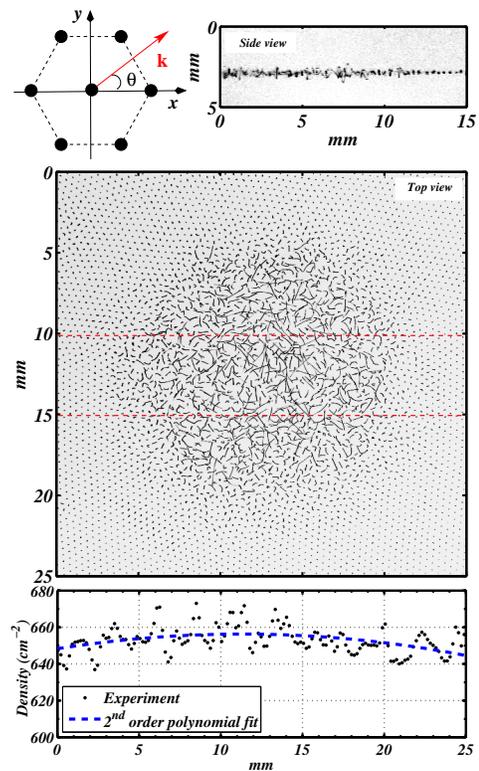}
\caption {\label{fig2} Example of the instability onset in a 2D plasma crystal. The shown example is for the experiment performed
at a rf power of 20~W with particles of $8.77~\mu$m diameter. The pressure of argon was ramped down in steps of 0.02~Pa (every
$\simeq300$~s), the instability was triggered at 0.90~Pa. The particle trajectories (superposition of consecutive video
frames during 0.04~s) are plotted about 50~s after the critical pressure decrease, the upper and middle panels demonstrate
the side and top views, respectively. The lower panel shows the profile of the particle density in the monolayer measured
{\it before} the instability onset (within the stripe marked in the middle panel by two horizontal dashed lines).
In the upper left corner the elementary cell of the hexagonal lattice is depicted with the frame of reference chosen in this
paper, due to the lattice symmetry we consider the wave vector ${\bf k}$ at $0^{\circ}\leq\theta\leq 30^{\circ}$.}
\end{figure}

One of two different discharge parameters was used as the control parameter for the melting: We either gradually decreased
the rf power at a constant gas pressure (which lowers the resonance frequency of the vertical confinement $f_{\rm v}$, see
Ref.~\cite{Steinberg01}), or we decreased the pressure at a constant rf power (which naturally lowers neutral gas friction,
but also makes the radial confinement stronger and hence the interparticle distance $a$ smaller) \cite{footnote_add}. In
either case, the melting developed in the same way, as illustrated in Fig.~\ref{fig2} (see also supplemental movie S1
\cite{suppl}): Upon decrease of the control parameter below a certain threshold the instability was triggered in the central
part of the crystal (i.e., where the particle density was the highest) and then expanded outwards. The instability always
set in simultaneously in the horizontal and vertical directions. At the initial stage it was accompanied by an exponential
growth of the particle kinetic energy, eventually resulting in the total disruption of the crystal in the unstable zone.
Upon increase of the control parameter back above the threshold (usually, with a hysteresis) the system always
recrystallized. It is noteworthy that if the rf power was sufficiently high and/or the particle density in the monolayer was
low enough the crystal remained stable at practically any pressure (e.g., at a rf power of 20 W, particle density of
160~cm$^{-2}$, and argon pressure down to $p\simeq0.4$~Pa -- the lowest value technically achievable with our setup).

The observed behavior fully agrees with the qualitative picture predicted by the theory of mode-coupling instability
\cite{Ivlev01,Ivlev03}: It can only set in when the out-of-plane and in-plane wave modes intersect, which is only possible
when $f_{\rm v}$ is sufficiently low or/and the particle density in the monolayer ($\propto a^{-2}$) is high enough. The
intersection should be accompanied by the simultaneous excitation of the vertical and horizontal oscillations. On the other
hand, the instability should be suppressed by sufficiently high gas friction, i.e., by increasing the gas pressure.

The chief result of our paper is the {\it ultimate proof} that the observed melting is indeed triggered by the mode-coupling
instability. The unique fingerprint characterizing the instability \cite{Ivlev01,Ivlev03} is the crossing of the
longitudinal in-plane and transverse out-of-plane DL branches, which results in the (wake-induced) resonance coupling
between them and the emergence of an unstable {\it hybrid mode} operating in a certain proximity of the crossing. This was
directly proved by measuring the dispersion relations of all three principal DL modes. For this purpose, we used the
technique by Nunomura {\it et al.} \cite{Nunomura02} of reconstructing the in-plane dispersion relations from thermal
fluctuations of particle velocities in the horizontal plane. In addition, we implemented a novel method of particle imaging
\cite{Couedel09} which allowed us to measure simultaneously the vertical fluctuations and hence to retrieve also the
out-of-plane mode. The essential advantage of this technique is that it does not involve any intermediate (model)
assumptions about the interparticle interactions, and therefore yields ``true'' dispersion relations (or ``fluctuation
spectra'') which can be obtained for arbitrary experimental conditions (that just allow the individual particle tracking).

\begin{figure}
\includegraphics[width=8.65cm,clip=]{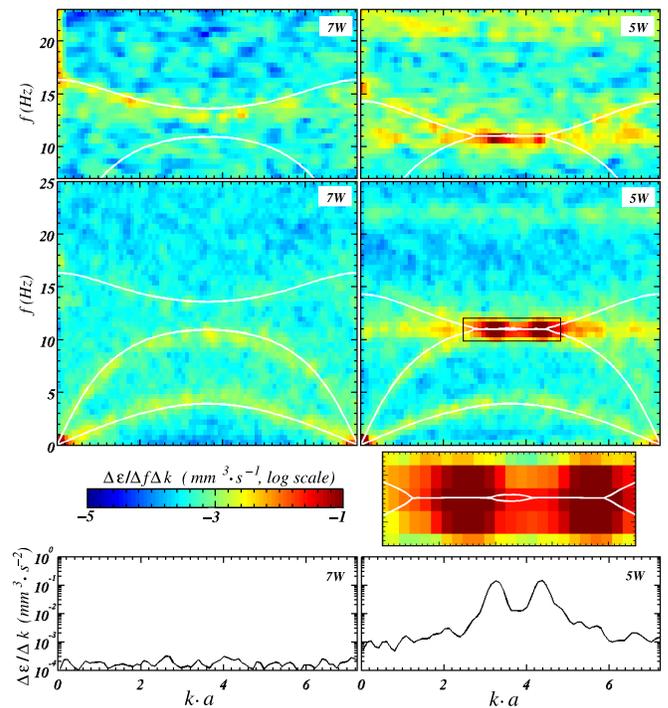}
\caption {\label{fig3} Evolution of the DL dispersion relations near the instability onset. The dispersions relations
(``fluctuation spectra'') were measured in the experiment performed at argon pressure of 0.76~Pa with particles of $9.15~\mu$m
diameter, at a rf power of 7 W (left column, stable regime) and 5 W (right column, the instability onset). The logarithmic
color coding is used for the ``fluctuation intensity'' $\Delta\epsilon/\Delta k\Delta f$ (energy density, for unit mass,
per unit wave number and frequency): Yellow ridges represent stable wave modes, the unstable hybrid mode are the dark-red
``hot spots''; the upper panel shows the measured out-of-plane mode, the second-from-the-top panel depicts the pair
of in-plane modes. The solid lines in both panels are theoretical results \cite{Zhdanov09} (see text for details). The wave
number $k$ is normalized by the inverse interparticle distance $a^{-1}$ (determined from the first peak of the pair correlation
function; the uncertainty of $a$ was $\simeq4\%$ due to density variations, see e.g., the lower panel of Fig.~\ref{fig2}). The shown range of $k$ corresponds to the first two
Brillouin zones at $\theta=30^{\circ}$ (branches are symmetric with respect to the border between the two zones). Below the
fluctuation spectra the magnified hybrid mode in the unstable regime (marked by a rectangular box) is shown. The lower
panel illustrates the distribution of the ``fluctuation energy'' $\Delta\epsilon/\Delta k$ at fixed frequency: In the stable
case the frequency ($\simeq12.5$~Hz) is chosen in the middle between the longitudinal in-plane branch and the out-of-plane
branch (measured at the border between the Brillouin zones), for the instability onset the frequency ($\simeq11$~Hz)
corresponds to the hybrid mode.}
\end{figure}

In Fig.~\ref{fig3} we plotted the fluctuation spectra for the out-of-plane mode (upper panel) separately from those for the
in-plane modes (second-from-the-top panel), because the imaging method implemented for the simultaneous tracking of the
vertical and horizontal fluctuations yields different background intensities \cite{Couedel09}. As regards the theoretical
curves, in each panel we plotted all principle modes that fit the frequency range shown.

Figure~\ref{fig3} summarizes results of our analysis. In the stable regime (left column) the wave modes are well separated
and the fluctuation intensity is evenly distributed over the branches. In contrast, the unstable regime (right column) is
characterized by the apparent intersection between the out-of-plane and (one of) in-plane branches. The intersection occurs
in the vicinity of the first Brillouin zone boundary, resulting in the anomalous energy release (dark-red ``hot spots'').

The qualitative effects and features of the mode-coupling instability mentioned above were originally predicted by the
theory for a 1D string \cite{Ivlev01,Ivlev03}, and they all remain unaltered also in 2D case. For rigorous quantitative
comparison with the experiments we employed the theoretical results for 2D crystals obtained recently by Zhdanov {\it et
al.} \cite{Zhdanov09}. The theoretical branches are shown in Fig.~\ref{fig3} by solid lines: The upper curve represents the
optical out-of-plane mode, the middle and lower curves are for the pair of acoustic in-plane modes (longitudinal and
transverse, respectively). These curves were calculated for the normalized dipole moment of the wake $\tilde q\tilde\delta$
(see Ref.~\cite{Zhdanov09}) equal to 0.1. We assumed the Yukawa interparticle interaction, with the parameters (viz., the
particle charge and the effective screening length) deduced from the low-$k$ part of in-plane fluctuation spectra (using the
method of Ref.~\cite{Nunomura02}).

Figure~\ref{fig3} shows that the theory yields excellent agreement with the experiment, both for the stable and unstable
regimes. Especially important is that in the unstable regime (right column) the {\it predicted} unstable hybrid mode
perfectly coincides with the locations of ``hot spots'' in the measured fluctuation spectra. Even subtle effects, such as
the splitting of ``hot spots'' at the border between the Brillouin zones is very well resolved (see the distribution of the
fluctuation energy in the lower panel, and also the magnified hybrid mode) \cite{footnote3}.

The fluctuation spectra shown in Fig.~\ref{fig3} were plotted for the wave vector ${\bf k}$ at $\theta=30^{\circ}$ (see
Fig.~\ref{fig2}), because the theory predicts this orientation to be the most unstable (see Fig. 4 of Ref.
\cite{Zhdanov09}). In order to further highlight excellent comprehensive agreement between experiments and theory, we
plotted additional spectra for different angles (see supplemental figure S2 \cite{suppl}). They clearly show that at
$\theta=30^{\circ}$ the fluctuation intensity of the ``hot spots'' is indeed the highest.


Note that rapid energy growth after the instability onset results in various nonlinear phenomena which can be easily
identified in the right column of Fig.~\ref{fig3}. The most prominent effects are the energy cascades from the ``hot spots''
along the $k$-axis (bright red-yellow stripe in the second-from-the-top panel, see also the energy profile in the lower
panel), and the generation of secondary harmonics at the double frequency of the hybrid mode (yellow bands at $\simeq22$~Hz
in the fluctuation spectra).

In conclusion, one should point out that any instability operating in 2D plasma crystals and associated with collective
processes \cite{footnote4} should be accompanied by anomalies in the fluctuation spectra. Therefore, the method of
simultaneous measurement of all three principal DL modes promises to be very fruitful for further investigation of the
melting scenarios. Especially important would be a systematic search for other possible resonant processes that can extend
or alter the mechanism reported in this Letter.

\newpage




\begin{thebibliography}{}
\bibitem{footnote1} For 2D plasma crystals, the ``trivial'' origin of the non-Hamiltonial behavior associated with neutral
    gas friction can often be neglected, since the corresponding damping rate is typically one-to-two orders of magnitude
    smaller that the Einstein frequency \cite{Morfill09}.
\bibitem{footnote_add} Changing either of these discharge parameters simultaneously affects all plasma
    characteristics (e.g., plasma density, particle charge, etc.). However, these parameters appear to be the most
    efficient in triggering the melting.
\bibitem{footnote3} Structure of the hybrid mode is also shown in Fig. 3 of Ref. \cite{Zhdanov09}; note the difference
    between the reference frames used in \cite{Zhdanov09} and in our paper.
\bibitem{footnote4} Here we leave aside the discussion of ``localized'' mechanisms of melting, e.g., due to
    defects.
\end{thebibliography}
\end{document}